# Galvanic intercalation of molecular cations into van der Waals materials

Daniel Tezze[1,2], Covadonga Álvarez[1,2], Daniel Margineda[1], Mohammad Furqan,[3,4] José Manuel Pereira[1,2], Umer Ahsan[5], Vlastimil Mazanek[5], Yogesh Kumar Maurya[6], Aurelio Mateo-Alonso[6,7], Frederik M. Schiller[8,9], Fèlix Casanova[1,7], Samuel Mañas-Valero[10], Eugenio Coronado[10], Iván Rivilla[7,9], Zdenek Sofer[5], Beatriz Martín-García[1,7], Maider Ormaza[2], Raul Arenal,[3,4,11] Luis E. Hueso[1,7], Marco Gobbi[7,8*]

[1] CIC nanoGUNE BRTA, Donostia-San Sebastian, 20018, Spain

[2] Departamento de Polímeros y Materiales Avanzados: Física, Química y Tecnología, Facultad de Químicas (UPV/EHU), Apartado 1072, 20080, San Sebastián, Spain

[3] Laboratorio de Microscopias Avanzadas (LMA), Universidad de Zaragoza, 50018 Zaragoza, Spain

[4] Instituto de Nanociencia y Materiales de Aragon (INMA), CSIC-Universidad de Zaragoza, 50018 Zaragoza, Spain

[5] Dept. of Inorganic Chemistry, University of Chemistry and Technology Prague, Technicka 5, 166 28 Prague 6, Czech Republic

[6] POLYMAT, University of the Basque Country UPV/EHU San Sebastian, 20018, Spain

[7] IKERBASQUE, Basque Foundation for Science, 48009 Bilbao, Spain

[8] Materials Physics Center (CFM-MPC) Centro Mixto CSIC-UPV/EHU, San Sebastian, 20018, Spain

[9] Donostia International Physics Center (DIPC), San Sebastián, 20018, Spain

[10] Instituto de Ciencia Molecular (ICMol), Universitat de València, 46980 Paterna, Spain

[11] ARAID Foundation, 50018 Zaragoza, Spain

[*]Corresponding authors:

marco.gobbi@ehu.eus

**The intercalation of molecular species between the layers of van der Waals (vdW) crystals is a powerful approach to combine the remarkable physical properties of vdW materials with the chemical versatility of organic molecules. However, the full transformative potential of molecular intercalation remains underexplored, largely due to the lack of simple, broadly applicable methods that preserve high crystalline quality down to the few-layer limit. Here, we introduce a simple galvanic approach to intercalate different molecules into various vdW materials under ambient conditions, leveraging the low reduction potential of selected metals. We employ our method, which is particularly well-suited for the in-situ intercalation of few-layer-thick crystals, to intercalate nine vdW materials, including magnets and superconductors, with molecules ranging from conventional alkylammonium ions to metallorganic and bio-inspired chiral cations. Notably, intercalation leads to an unprecedented transition from antiferromagnetic to ferrimagnetic ordering in $\alpha$-$RuCl_3$ and to a molecule-dependent enhancement of the superconducting transition in 2H-$TaS_2$. These results establish our approach as a versatile technique for engineering atomically thin quantum materials and heterostructures, unlocking the transformative effects of molecular intercalation.**

Intercalation – the insertion of guest species into the gaps of layered van der Waals (vdW) materials – has emerged as a powerful strategy for engineering hybrid systems with tailored functionalities. While alkali and alkaline-earth metals have historically been the most common guest species,[1] recent advances in molecular intercalation have unlocked novel opportunities to combine the electronic properties of vdW materials with the chemical versatility of organic molecules.[2–7] Molecule-intercalated vdW crystals form hybrid superlattices where molecular monolayers are embedded between vdW planes,[8] creating a tunable material platform to control electronic[8–12], magnetic[13–16] and optical[17] properties. Recent studies have shown how selecting molecules with specific functionalities, such as chirality[18–20] or magnetism[21,22], can induce novel phenomena in vdW materials.

Molecular intercalation is conventionally achieved through chemical or electrochemical methods[4,23–26]. Chemical intercalation involves immersing vdW crystals in a solvent containing target molecules, which spontaneously occupy the vdW gaps under equilibrium conditions[27,28]. However, this process is constrained by the stringent alignment between frontier energy levels of host and guest, greatly limiting the choice of vdW materials and molecules[28,29]. Besides, harsh conditions such as high temperature, pressure, or prolonged reaction time are required, which often compromise the material integrity, in particular when few-nm-thick crystals are processed[30,9]. Electrochemical intercalation employs an external potential to trigger the intercalation process in a broader class of vdW materials[8,24,29,31,32]. Despite its versatility, the non-equilibrium nature of this method makes it inherently aggressive and difficult to control,[14,33] often yielding intercalated compounds with lower structural quality than pristine materials.[10,14] While traditionally used for bulk crystals[10,11,14,34–36], electrochemical intercalation has recently been adapted for few-nm-thick flakes contacted by micrometric electrodes[8,37–39]. However, this in-situ intercalation requires a complex device fabrication and a highly controlled environment to minimize side reactions, host material exfoliation and degradation of electrical contacts[8,38].

Here, we introduce a simple galvanic intercalation strategy tailored for few-nanometer-thick flakes, enabling the intercalation of diverse vdW materials with molecular cations under mild conditions. To demonstrate its broad applicability, we tested the intercalation of multiple vdW hosts, including magnets and superconductors, with a variety of molecular guest species, such as alkylammonium, chiral, and organometallic cations. We found that nine vdW materials can be intercalated with at least two different molecular species, yielding 49 novel organic-inorganic superlattices with structural quality comparable to the pristine crystals. To demonstrate the potential of our approach to tailor physical properties, we focus on superconducting 2H-$TaS_2$ and magnetic α-$RuCl_3$ as case studies. Intercalation induces a molecule-dependent enhancement of the superconducting critical temperature in 2H-$TaS_2$, reaching 4.7 K with chiral-molecule intercalation, surpassing even $TaS_2$ monolayers.[40] Additionally, α-$RuCl_3$ undergoes an unprecedented transition from antiferromagnetic to ferrimagnetic ordering upon intercalation with cobaltocenium. These results underscore the potential of our galvanic intercalation method to tailor materials structures with atomic-scale precision, hence unlocking emergent quantum states and functionalities.

## Results and discussion

### The molecular galvanic intercalation method

In our method, a vdW host material is electrically connected to a metal (M) with a low reduction potential, such as $In^0$, $Zn^0$ or $Mg^0$ (Fig. 1a). For example, the host can be a vdW flake transferred onto a gold-coated substrate, electrically in contact with $M^0$. Then, both host and M are immersed in a non-aqueous solution containing a salt of the target molecular guest ($G^+$). Under these conditions, the system behaves as a galvanic cell. The low-reduction-potential M acts as the anode, undergoing spontaneous oxidation ($M^0 \rightarrow M^{x+} + xe^-$, x = 1, 2, 3, …). The host serves as the cathode, being electrochemically reduced by electrons provided by the metal oxidation through the electrical connection (host + $e^- \rightarrow$ host$^-$). To preserve charge neutrality within the vdW material, molecular ions are introduced into the vdW gap, leading to the intercalation (host$^-$ + $G^+ \rightarrow$ G-host). This intercalation mechanism implies that for each $G^+$ introduced in the vdW gap, an electron is formally introduced in the band structure of the host, leading to a large charge carrier doping (on the order of $10^{14}$ charges/cm$^2$ in each layer).[41] While an analogous approach was employed to intercalate different vdW compounds with alkali and alkaline earth metals (e.g. $Li^+$)[42], it has not been explored to achieve molecular intercalation. We highlight that the whole process can be carried out in air, as $In^0$, $Zn^0$ or $Mg^0$ are stable and can be handled in air. Overall, this galvanic approach retains the simplicity of chemical intercalation, as the process occurs spontaneously upon immersing the crystal in solution, while significantly expanding its applicability to a broader range of materials.

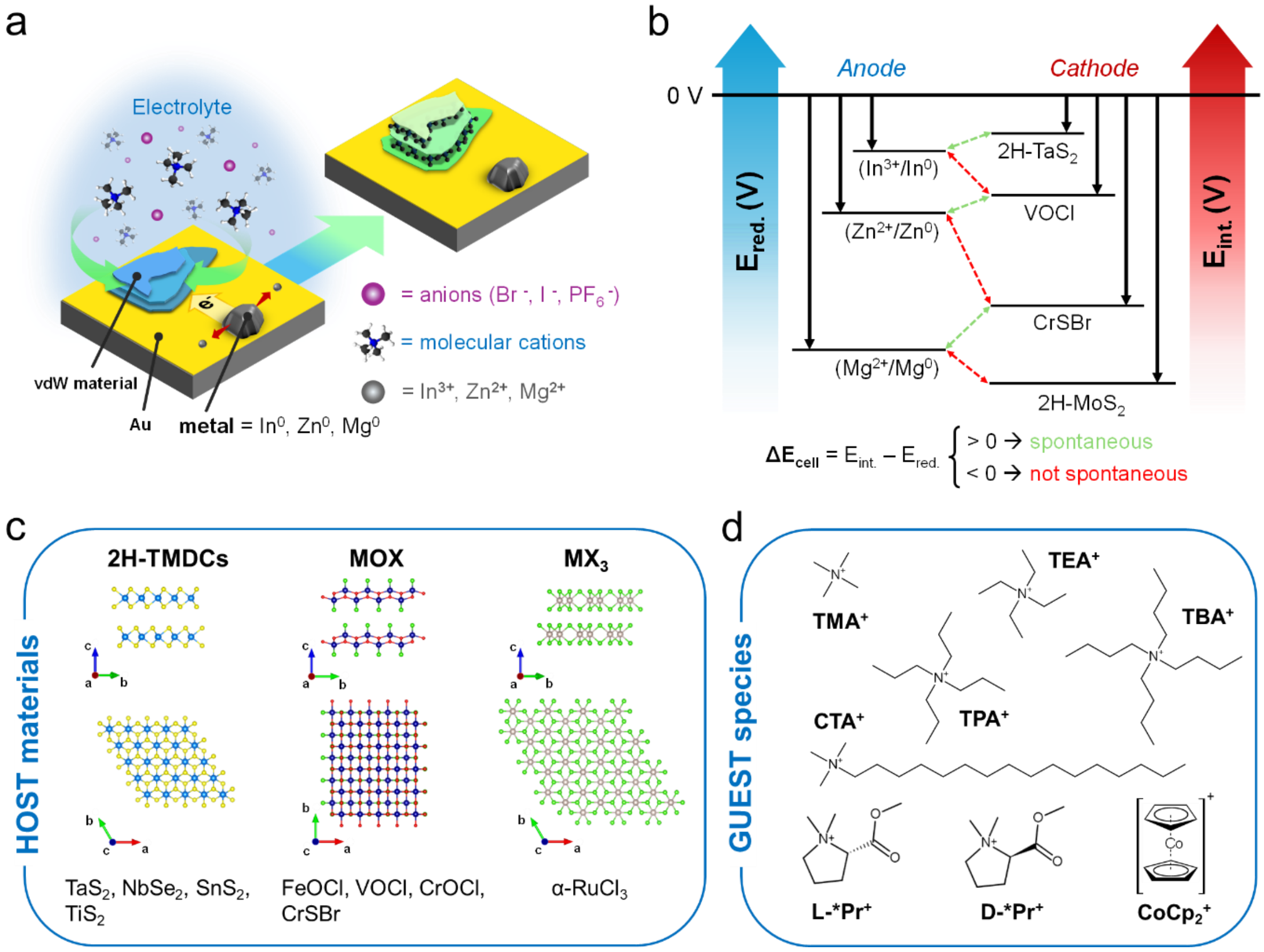

***Figure 1. Concept of Galvanic Intercalation: (a)*** *Sketch of the engineered galvanic cell for intercalating molecular cations into a few-layer thick vdW crystal (cathode), using a suitable metal ($In^0$, $Zn^0$, $Mg^0$) as the anode. When the substrate is immersed in a solution, spontaneous half-reactions generate an in-built electrical potential that drives the molecular intercalation.* ***(b)*** *The process occurs spontaneously if the reduction potential for intercalation $E_{int}$ of the vdW material lies above the reduction potential of the anodic metal ($E_{red.}$). For example, VOCl can be intercalated with $TMA^+$ ions in ACN (5 mM) using $Zn^0$ as the anodic metal, but not $In^0$, indicating that its $E_{int.}$ lies between the $E_{red.}$ of the two metals.* ***(c)*** *Layered vdW materials successfully intercalated via galvanic intercalation.* ***(d)*** *Molecular cations chosen as guest species for the hybrid superlattices in this work: alkylammonium ions ($TMA^+$, $TEA^+$, $TPA^+$, $TBA^+$, and $CTA^+$), the two enantiomers of a prolinium derivative (L –*$Pr^+$ and D –*$Pr^+$), and the organometallic cobaltocenium ion ($CoCp_2^+$).*

The intercalation process occurs spontaneously when the reduction potential of the anodic metal ($E_{red.}$) is lower than the reduction potential for the intercalation of a certain vdW material ($E_{int}$) (**Figure 1b**). $E_{int}$ depends on different factors, such as the electrolyte, its concentration and the solvent. Among the metals tested, $Mg^0$ has the lowest reduction potential, while $In^0$ has the highest under standard aqueous conditions. Our experiments, conducted with various vdW material/molecule combinations, confirm that this trend remains unchanged within the different galvanic cells tested in this study. Consequently, if a vdW material can be intercalated using $In^0$ as the anodic metal, it can also be intercalated with $Zn^0$ and $Mg^0$, as observed for 2H-$TaS_2$ and α-$RuCl_3$. Conversely, some materials, such as CrSBr, can be intercalated with $Mg^0$ but not with $Zn^0$ or $In^0$. Finally, certain compounds, like 2H-$MoS_2$, cannot be intercalated with any of the three metals tested, suggesting that their $E_{int.}$ lies below the $E_{red}$ of $Mg^0$ (Figure 1b).

The families of vdW compounds which we successfully intercalated using the galvanic approach include superconducting transition metal dichalcogenides[43], magnetic metal oxyhalides and trihalides[44] (**Figure 1 c**). Notably, most of these materials can be chemically intercalated with Lewis bases[28], whereas the intercalation of cations typically requires a more demanding electrochemical approach. As guest molecular cations, we focused on common alkylammonium ions that differ in size and shape ($TMA^+$, $TEA^+$, $TPA^+$, $TBA^+$, $CTA^+$), a chiral prolinium-derivative (left-hand L-*$Pr^+$ and right-handed D-*$Pr^+$) and cobaltocenium ($CoCp_2^+$), as depicted in **Figure 1d**. We successfully intercalated at least two of these molecules into each of the nine different vdW compounds tested, achieving a total of 49 organic-inorganic superlattices (Supplementary Table 1). Our method is versatile and adaptable for intercalating bulk crystals, multiple exfoliated flakes and individual few-nm-thick flakes.

**Molecular galvanic intercalation of bulk crystals**

We employed an experimental setup as the one shown in **Figure 2a** to intercalate bulk crystals of α-$RuCl_3$, with $CoCp_2^+$ ions and 2H-$TaS_2$ with $TMA^+$ ions. The galvanic intercalation process is accompanied by a spontaneous current at zero applied voltage, generated by the in-build electrical potential between the short-circuited cathode and the anode, which can be measured by an amperemeter connected to the external wire. **Figure 2b** shows the evolution of the galvanic current over time for the intercalation of α-$RuCl_3$ with $CoCp_2^+$ ions. Based on this data, we estimated the stoichiometric index x = 0.27 ± 0.01 for $(CoCp_2)_xRuCl_3$, averaging over different crystals (Supplementary Table 2).

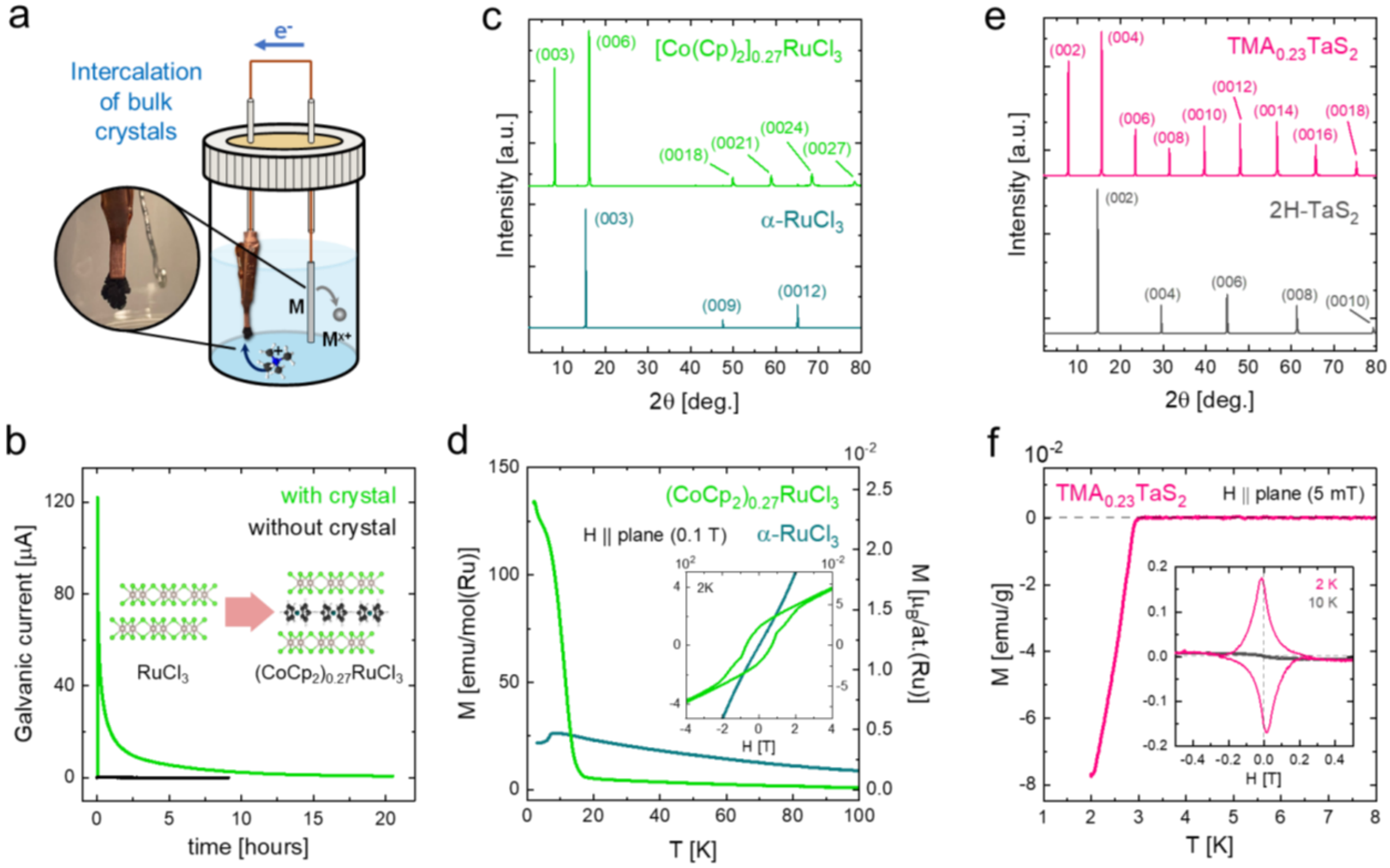

***Figure 2. Tuning magnetism and superconductivity in bulk crystals through galvanic molecular intercalation: (a)*** *Experimental setup for the galvanic molecular intercalation process of bulk crystals.* ***(b)*** *Galvanic current over time recorded at zero applied voltage during the intercalation of a α-$RuCl_3$ crystal ($m_p$ = 2.66 mg) with $Co(Cp)_2^+$ ions, using a 10 mM $CoCp_2PF_6$/ACN electrolyte (30 mL) and $Zn^0$ as the anode.* ***(c)*** *X-ray diffraction pattern acquired for a pristine α-$RuCl_3$ crystal and after the intercalation process with $CoCp_2^+$ ions.* ***(d)*** *Temperature dependence of the in-plane magnetization, M(T), of a pristine α-$RuCl_3$ crystal and a $(CoCp_2)_{0.27}RuCl_3$ crystal, recorded at H = 0.1 T. Inset: in-plane hysteresis loop recorded at 2K for pristine α-$RuCl_3$ crystal and $(CoCp_2)_{0.27}RuCl_3$ crystal.* ***(e)*** *X-ray diffraction pattern measured for a pristine 2H-$TaS_2$ crystal and after intercalation with $TMA^+$ ions, using a 5 mM TMAB/MeOH electrolyte (30 mL) and $In^0$ as the anode.* ***(f)*** *In-plane zero-field-cooled (ZFC) magnetization over temperature, M(T), of pristine and $TMA_{0.23}TaS_2$, recorded at H = 5 mT. Inset: Field dependence of the magnetization of $TMA_{0.23}TaS_2$ recorded at 2 K and 10 K.*

The intercalation of α-$RuCl_3$ with $Co(Cp)_2^+$ was confirmed by X-Ray diffraction (XRD). In **Figure 2c** we compare the XRD patterns of a bulk α-$RuCl_3$ crystal measured before and after galvanic intercalation. After intercalation, the XRD pattern of the crystal exhibits one set of very sharp (00l) peaks shifted to lower angles compared to the pristine material, indicating an increase in interlayer distance from 5.7(4)

Å to 10.9(3) Å (Δd = 5.19 Å), to accommodate $CoCp_2^+$. The pristine and intercalated crystals display diffraction peaks with similar full width at half maximum, revealing almost unaffected crystalline quality. X-ray photoelectron spectroscopy revealed the reduction of a fraction of Ru(III) to Ru(II) in α-$RuCl_3$, due to the charge transfer induced by intercalation (Supplementary Fig. 1).

The magnetic properties are also dramatically changed by intercalation. While pristine α-$RuCl_3$ exhibits antiferromagnetic behavior with a Neel temperature of ~ 7 K[45], $(CoCp_2)_{0.27}RuCl_3$ shows a spontaneous magnetization below ~ 13 K (**Figure 2d**). At 2 K, $(CoCp_2)_{0.27}RuCl_3$ displays a magnetic hysteresis characterized by a finite remanence and a large coercivity ($H_c$ ~ 7.0 kOe, inset in **Figure 2d**). Additional magnetic characterization of pristine and $CoCp_2^+$ intercalated α-$RuCl_3$ is presented in Supplementary Fig. 2. Based on this magnetic characterization, we conclude that $(CoCp_2)_{0.27}RuCl_3$ is a ferrimagnetic compound with in-plane anisotropy and enhanced transition temperature. Notably, long-range magnetic ordering was not observed in α-$RuCl_3$ intercalated with $TMA^+$ ions in a previous study[46], indicating an active role of the metallorganic $CoCp_2^+$ ions in the emergent magnetization, in agreement with recent works on $CoCp_2$-intercalated 2H-$SnS_2$[22].

Next, we focus on the effect of intercalation of $TMA^+$ on 2H-$TaS_2$, which is a superconductor with $T_c$ ~ 0.8 K in its pristine state.[40] XRD confirms the successful generation of an high-quality hybrid superlattice (**Figure 2e**). To characterize the superconductive nature of the intercalated crystal, we measured the temperature dependence of the zero-field-cooled magnetization (**Figure 2f**). The negative magnetization recorded below 2.95 K, signature of the Meissner effect, indicates the formation of superconducting correlation at an enhanced $T_c$. The magnetic field dependence of the magnetization measured at 2K presents a hysteresis, indicating that TMA-$TaS_2$ is a type-II superconductor (inset in **Figure 2f**)[47]. X-ray fluorescence measurements performed on galvanically intercalated α-$RuCl_3$ and 2H-$TaS_2$ bulk crystals detected no traces of Zn and In (Supplementary Fig. 3).

**Molecular galvanic intercalation of multiple few-nm-thick crystals**

The galvanic approach is well-suited for the intercalation of randomly distributed flakes transferred on a conducting substrate (see **Figure 3a** and Methods). In this configuration, the anodic metal is electrically connected to the substrate, and both are immersed in a solution of the chosen molecular salt. **Figure 3b** shows representative optical images of 2H-$TaS_2$ flakes after galvanic intercalation with $TMA^+$ and $TEA^+$ ions (5 mM, 1h, $In^0$ anode, RT). Notably each $TMA^+$-intercalated flake displays a uniform color, indicating complete intercalation. In contrast, $TEA^+$-intercalated flakes exhibit variations in optical contrast between their inner and outer regions. This contrast arises because intercalation is a topotactic process which proceeds from the flake's perimeter towards the center[15,48]. For some species, such as $TEA^+$ and $CoCp_2^+$ ions, the intercalation proceeds slowly enough that it can be halted before completion, resulting in partially intercalated flakes. However, given sufficient time, complete intercalation can be achieved even for these molecules, yielding homogeneously intercalated flakes.

The galvanic intercalation approach is straightforward and closely resembles chemical intercalation, except for the presence of the anodic metal in electrical contact with the flakes. To compare the two approaches, we present the XRD patterns of 2H-$TaS_2$ intercalated with $TMA^+$ using either the chemical or the galvanic method (**Figure 3c**). Whereas a successful intercalation is achieved after 1h at room temperature using the galvanic method, chemical intercalation requires significantly higher temperatures (60 °C in MeOH) and a much longer duration (~ 1 day) to be effective. Notably, the galvanic approach exhibits higher crystalline quality, as indicated by the narrower XRD peaks. We point out that chemical intercalation using salts as the reagent is only possible for a limited number of van der Waals materials, including 2H-$TaS_2$, FeOCl, and α-$RuCl_3$. Our galvanic method significantly expands the range of materials that can be intercalated using salts without an electrochemical approach.

Given the simplicity and efficiency of the galvanic intercalation of flakes, we used this method to systematically evaluate the compatibility of various vdW materials and molecules for generating a library of hybrid superlattices. All the molecules displayed in **Figure 1d** were successfully intercalated in 2H-$TaS_2$, FeOCl and α-$RuCl_3$, as confirmed by XRD measurements (Supplementary Figs. 4-6). Moreover, other six vdW compounds could be intercalated with at least two molecules (Supplementary Figs 7-12). In **Figure 3d** we report a representative set of six XRD patterns taken from hybrid superlattices of 2H-$TaS_2$ compared with the pristine one. All the intercalated compounds display high crystalline quality, as indicated by the sharp and well-defined XRD peaks. Additionally, we tested an ion-exchange approach to substitute $TMA^+$ with $CoCp_2^+$ ions[14] in galvanically pre-intercalated 2H-$SnS_2$ and CrOCl, further increasing the number of guest molecules compatible with our approach (Supplementary Fig. 13).

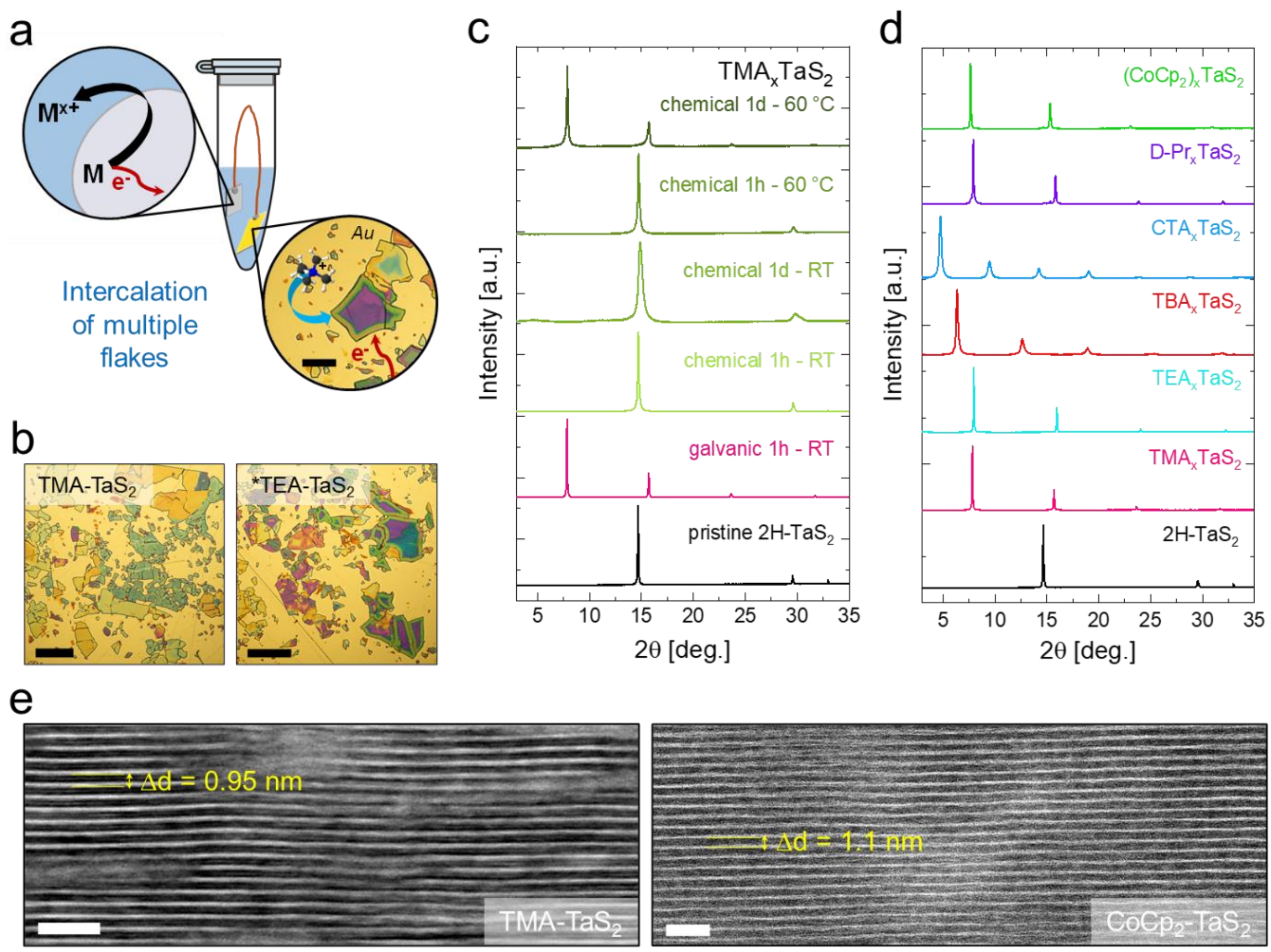

***Figure 3. Galvanic molecular intercalation of different molecules in multiple exfoliated flakes: (a)*** *Scheme of the setup used for the galvanic intercalation of multiple flakes exfoliated and transferred onto an Au thin film coated* $SiO_2$ *substrate. Scale bar: 10 µm.* ***(b)*** *Optical images of 2H-*$TaS_2$ *flakes intercalated with* $TMA^+$ *ions (Scale bar: 20 µm) and partially intercalated with* $TEA^+$ *ions (bar length: 50 µm). [TMAB] = [TEAB] = 5 mM in MeOH,* $In^0$ *anode, 1h.* ***(c)*** *Comparison between the XRD patterns measured for 2H-*$TaS_2$ *flakes intercalated with* $TMA^+$ *ions using either the galvanic or the conventional chemical approach.* ***(d)*** *X-ray diffraction patterns measured for multiple pristine and intercalated 2H-*$TaS_2$ *exfoliated flakes, with six molecular ions.* ***(e)*** *Cross-sectional scanning transmission electron microscopy images of 2H-*$TaS_2$ *flakes intercalated with* $TMA^+$ *and* $Co(Cp)_2^+$ *ions, showing the high crystallinity of the organic-inorganic superlattices. For TMA-*$TaS_2$*: [TMAB] = 5 mM in MeOH,* $In^0$ *anode, 1h. For* $CoCp_2$*-*$TaS_2$*: [*$CoCp_2PF_6$*] = 10 mM and [TMAB] = 1 mM in ACN,* $In^0$ *anode, 2h. Scale bar: 5 nm.*

The high quality of the hybrid superlattices is further confirmed through Scanning Transmission Electron Microscopy (STEM). **Figure 3e and f** present focused ion beam (FIB) prepared cross-sectional High-Angle Annular Dark-Field STEM (HAADF-STEM) images of 2H-$TaS_2$ flakes intercalated with $TMA^+$ and $CoCp_2^+$ ions, respectively. Both crystals exhibit a well-ordered layered structure with an interlayer distance of d = 0.95 nm in case of $TMA^+$ and d = 1.1 nm for $CoCp_2^+$, in good agreement with XRD data. Remarkably, the ordered regions imaged in **Figures 3e** and **f** extend over approximately 50 nm × 15 nm, indicating that intercalation preserves long-range structural order. TEM images displaying even wider areas (Supplementary Fig. 14) further demonstrate that intercalation homogeneously extends across the entire flake thickness. Higher-resolution STEM images (Supplementary Fig. 14) confirm the atomic-scale structural integrity of the intercalated crystals and the high quality of the hybrid interfaces. Energy-Dispersive X-ray Spectroscopy (EDS) STEM has also been performed to confirm the composition of these systems (Supplementary Fig. 15).

Another powerful tool to characterize the intercalation at the single-flake level is micro-Raman spectroscopy, as the phonon modes are profoundly modified by intercalation. As an example, we focus on $TaS_2$. The Raman spectrum of pristine 2H-$TaS_2$ presents three characteristic Raman-active peaks, corresponding to the two-phonon (186 $cm^{-1}$), $E_{2g}$ (286 $cm^{-1}$), and $A_{1g}$ (402 $cm^{-1}$) modes (**Figure 4a**).[49] After intercalation, the spectrum undergoes dramatic changes that are largely independent of the molecular guest. Specifically, the $^1A_{1g}$ peak shifts considerably to 394 $cm^{-1}$, the two-phonon mode slightly moves to higher wavenumbers, and the $^2E_{2g}$ peak is suppressed (**Figure 4a**). Additionally, intercalation leads to an overall increase in peak intensity.

Thanks to these pronounced differences, micro-Raman spectroscopy can be used to probe intercalation with micrometric spatial resolution. For example, **Figure 4b** shows a representative optical image of a partially intercalated flake, which was intercalated with $TEA^+$ ions for 1 hour using $In^0$ as the anode. A map of the $^1A_{1g}$ peak intensity shows a clear enhancement of the Raman signal in the region close to the flake's edge, confirming the partial intercalation and the presence of a sharp interface between the pristine and intercalated phases. Notably, Raman spectra of both thick and few-layer flakes exhibit analogous features, indicating that the intercalation process is largely independent of flake thickness (**Figure 4c**).

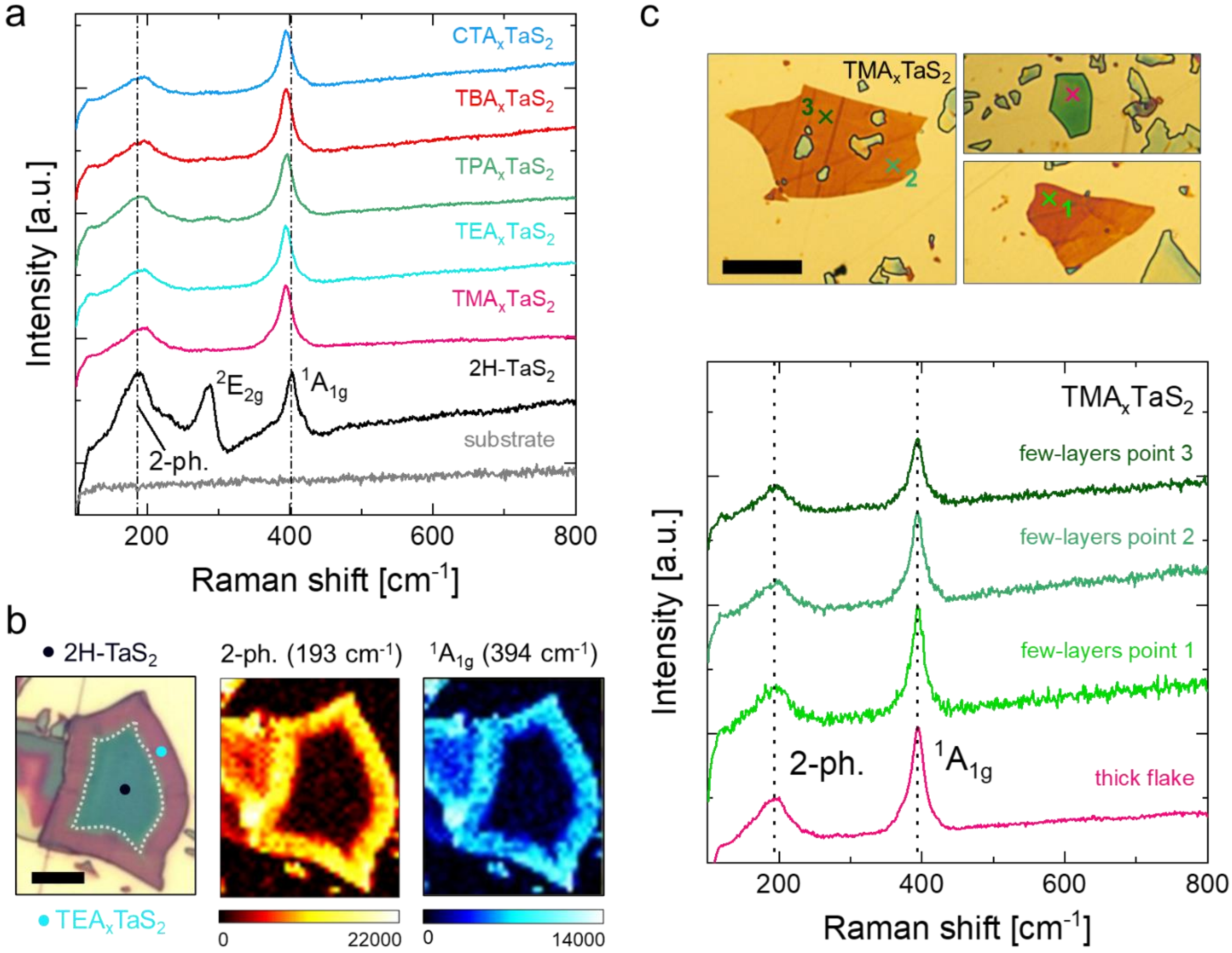

***Figure 4: Raman spectroscopy characterization of intercalated 2H-TaS2 flakes. (a)*** *Micro-Raman spectra measured for pristine 2H-TaS$_2$ and TMA$^+$,TEA$^+$, TPA$^+$, TBA$^+$ and CTA$^+$ intercalated 2H-TaS$_2$ flakes. Spectra were recorded in ambient conditions, using a 532 nm laser.* ***(b)*** *Optical image (left) and corresponding Raman maps of the intensity of the two-phonon mode and the $^1A_{1g}$ mode (right) of a partially intercalated 2H-TaS$_2$ flake with TEA$^+$ ions. [TEAB] = 5 mM in MeOH, In$^0$ anode, 1h. Scale bar: 10 µm.* ***(c)*** *Optical image (top) and micro-Raman spectra (bottom) of different TMA$^+$-intercalated 2H-TaS$_2$ flakes, including few-layer-thick crystals. [TMAB] = 5 mM in MeOH, In$^0$ anode, 1h). Scale bar: 10 µm.*

**In-device Intercalation of an individual van der Waals thin crystal**

Galvanic molecular intercalation can also be successfully performed on preselected individual flakes integrated in a device, which represents an ideal testbed for measuring transport properties. To explore this approach, we have focused on 2H-TaS$_2$ intercalated with different molecules. Intercalating single flakes involves transferring the target flake onto pre-patterned metallic electrodes, connecting the anodic metal to one of the electrodes, and covering the device with a drop of the solution of the target molecular salt, which closes the circuit of the galvanic cell-on-a-chip (**Figure 5a).** The homogeneous intercalation of a flake, which changes color during intercalation, is confirmed by Raman spectroscopy (Supplementary Video 1 and Supplementary Fig. 16).

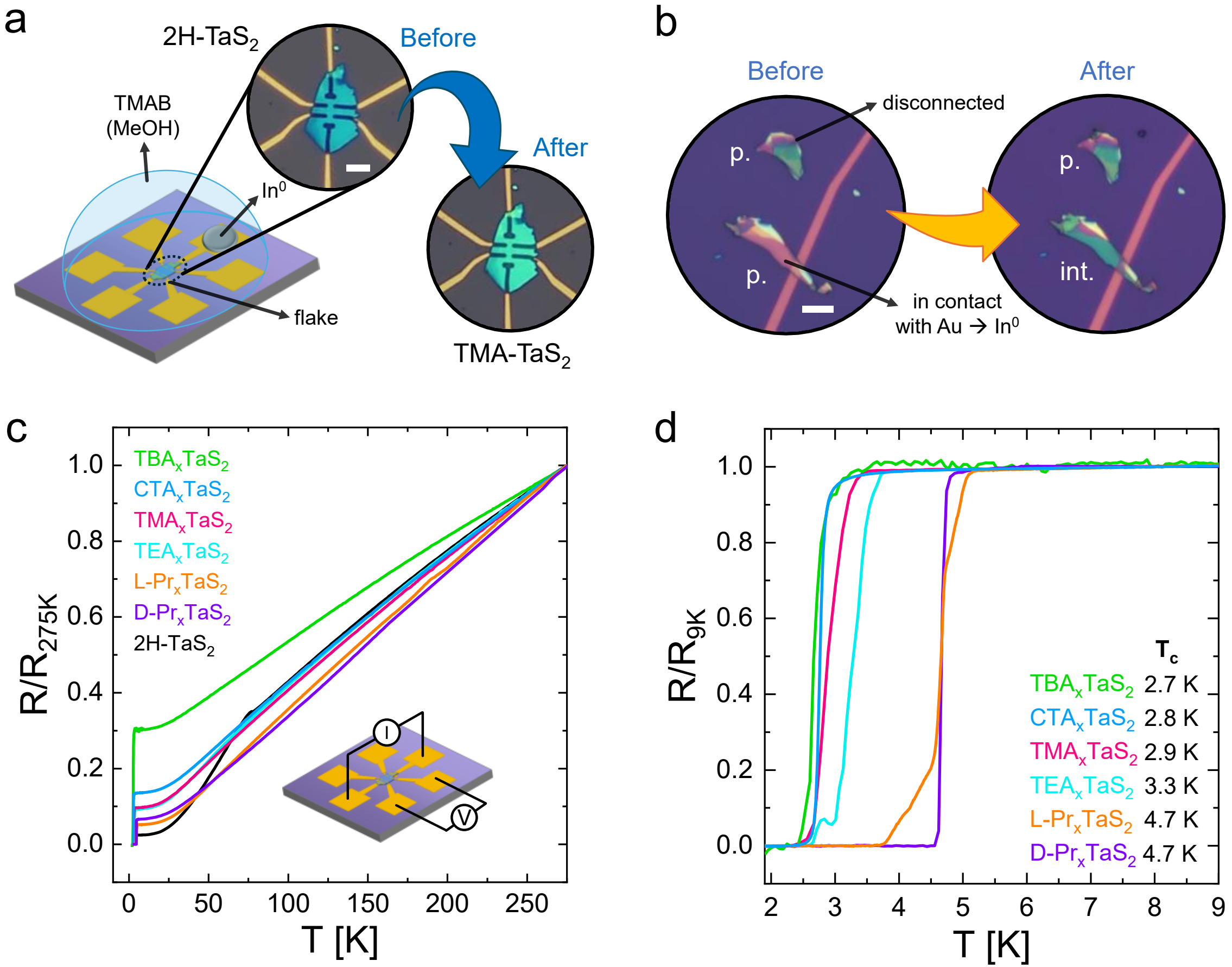

***Figure 5. Galvanic molecular intercalation of different molecules in single-flake devices: (a)*** *Optical image of a single flake of 2H-$TaS_2$ stamped on $SiO_2$/Si substrate with pre-patterned Au contacts before (left) and after (right) the intercalation process. The color change, due to the change in the thickness of the flake, is the first indication of intercalation success.* ***(b)*** *Optical images of two flakes acquired before and after adding a drop of the electrolyte onto the substrate. Only the flake connected to $In^0$ through the Au electrode is intercalated, as confirmed by the color change.* ***(c-d)*** *Temperature dependence of the resistance of 2H-$TaS_2$ flakes intercalated with the different molecules in the ranges 1.9 K – 275 K and 1.9 K – 9 K.*

This single-flake intercalation also provides additional experimental evidence of the efficiency of the galvanic approach. **Figure 5b** shows an area of the substrate with an Au electrode connected to an $In^0$ piece and two 2H-$TaS_2$ flakes fortuitously close to each other, with only one in contact with Au. After applying a drop of the target electrolyte, a clear color change is observed for the flake touching the Au electrode, but not for the other, indicating that only the flake electrically connected to In is effectively intercalated.

Next, we focus on the superconducting tunability of 2H-$TaS_2$ vdW materials. **Figure 5c** shows the temperature dependence of the resistance R(T) of 2H-$TaS_2$ intercalated with six different organic cations in the temperature range 1.9 K – 275 K. In all cases, a metallic behavior is recorded after intercalation. At low temperatures, all intercalates exhibit a transition to a superconducting state (**Figure 5d**). The transition is remarkably sharp, indicating a homogeneous superconducting state. We note that pristine 2H-$TaS_2$ flakes with a number of layers comparable to that used in this experiment have a critical temperature $T_c$ of approximately 0.8 K,[40] not accessible in our setup. The intercalation of $TMA^+$, $TEA^+$, $TBA^+$ and $CTA^+$, which are chemically similar alkylammonium cations but in different sizes, results in $T_c$ values between 2.7 K and 3.3 K. These $T_c$ are higher than that of bulk 2H-$TaS_2$ crystal and comparable to 1H-$TaS_2$ monolayer's, suggesting that these alkylammonium compounds act as spacers, increasing the interlayer separation in $TaS_2$, and providing monolayer behavior to the few-nm-thick flakes. Remarkably, a $T_c$ of 4.7 K is recorded for both L/D-*$Pr^+$-$TaS_2$ intercalates, higher than that of 2H-$TaS_2$ monolayer[40] and close to the highest $T_c$ reported for molecule-intercalated 2H-$TaS_2$[9,50]. Beside the alkylammonium moiety, L/D-*$Pr^+$ possesses a polar C=O group, which may hybridize or interact with dipolar interaction with 2H-$TaS_2$, possibly being responsible for the anomalously high $T_c$.

**Conclusion**

Our results demonstrate that galvanic molecular intercalation is a versatile and effective method to incorporate diverse molecular cations into a wide range of vdW materials, from bulk crystals to individual few-layers flakes. The numerous materials synthesized in this study, many of which remain to be fully characterized in detail, are anticipated to exhibit novel magnetic and superconductive properties. Tuning intercalation conditions – such as testing different electrolytes, anodic metals, concentrations, and temperatures – will expand the applicability of the galvanic approach to a broader range of host-guest combinations, potentially including other functional molecules specifically designed for quantum[51] or neuromorphic computing. Additionally, our approach enables the creation of lateral heterostructures between intercalated and pristine regions in the same crystal, a key requirement for fabricating advanced electronic devices. It also allows for the selective intercalation of target flakes integrated in vdW heterostructures, potentially unlocking new functionalities in vdW materials and devices.

## Methods

Materials: *Van der Waals crystals* – 2H-$TaS_2$ crystals were purchased from HQ Graphene. 2H-$NbSe_2$, 2H-$TiS_2$, 2H-$SnS_2$, FeOCl, VOCl, CrOCl, α-$RuCl_3$ crystals were made by chemical vapor transport (CVT). Synthesis protocols are reported in Supplementary Methods 1. CrSBr crystals were synthesized by CVT and characterized by powder and crystal X-ray diffraction, energy dispersive X-ray analysis, high-resolution transmission electron microscopy, superconducting quantum interference device magnetometry and temperature-dependent single crystal diffraction, as reported previously[52].

*Anodic metals* – Indium wire (purity 99.99 %) was purchased from CMR-Direct, Zinc (purity 99,9%) were purchased from G. L. Y. (by Amazon S.p.A.) and Magnesium (purity 99.95%) from S. X. Keji Co. Ltd (by Amazon S.p.A.). Platinum plates were obtained by pressing metallic pellets (purity 99.99%, Kurt J. Lesker). Before any usage, $Zn^0$ and $Mg^0$ plates are polished with sandpaper and rinsed with iso-propanol. Platinum plates are polished with sandpaper and sonicated for two minutes in acetone and then in isopropanol for surface cleaning.

*Organic Salts* – Tetramethylammonium bromide (TMAB, purity ≥ 98 %), tetraethylammonium bromide (TEAB, purity = 98%), tetrapropylammonium bromide (TPAB, purity = 98 %), tetrabutylammonium bromide (TBAB, purity ≥ 98 %), bis-(cyclopentadienyl)cobalt(III) hexafluorophosphate ($Co(Cp)_2PF_6$, purity 98%) and and cetyltrimethylammonium bromide (CTAB, purity ≥ 99%) was purchased from Across Organics. Before usage, salts were dehydrated at 100° C in vacuum (~ 1 mbar) overnight, expect for Bis(cyclopentadienyl)cobalt(III) iodide ($CoCp_2I$) and (S)- and (R)-2-(methoxycarbonyl)-1,1-dimethylpyrrolidin-1-ium iodide (L-*PrI and D-*PrI) ,which were only vacuum dried (~ 1 mbar) overnight after their synthesis.

$CoCp_2I$ was obtained following the reported procedure[53]. L-*PrI and D-*PrI were synthesized and characterized as reported in Supplementary Methods 2.

*Other reagents* – D- and L-Proline Methyl Ester Hydrochloride (L-Pro-OMe HCl, purity 98 %), and cyclopentadienyl)cobalt(II) ($CoCp_2$, MQ100) were purchased from Aldrich. Iodomethane (MeI, purity 99%) was purchased from Merck. Iodine ($I_2$) was purchased at vwr (purity: 99.5 %). All the reagents were used without any further purification.

*Solvents* – Methanol (anhydrous, purity 99.8%), N-N-Dimethylformamide (DMF, purity ≥ 99.8 %), Dimethyl sulfoxide (DMSO, purity ≥ 99.9 %), Propylene carbonate (PC, anhydrous, purity 99.7%), benzene (anhydrous, purity 99.8 %) were purchased at Sigma-Aldrich. Acetonitrile (ACN, anhydrous, purity 99.9 %, for electrolytes) was purchased at Scharlab. Ethyl acetate (99,5 %) and acetonitrile (HPLC grade, for $CoCp_2I$ synthesis) were purchased at Fisher. All the solvents were used without any further purification.

Galvanic Intercalation: *Bulk crystals* – For intercalating bulk crystals, a rudimental two-electrode cell is employed. For intercalation with an $In^0$ as the anode, a bulk crystal is clamped with Cu crocodile clip and used as a cathode (Figure 2a). When $Zn^0$ or $Mg^0$ are required, the bulk crystal is fixed to a platinum plate by means of $In^0$ strips (see Supplementary Figure 17). The crystal is electrically connected to the anodic metal of choice through a copper wire. A source-measure unit operating as amperemeter (Keithley 2635) is placed in between the two electrodes to measure the current flow versus time. The two electrodes are immersed in the electrolyte of choice. The intercalation starts spontaneously when the electrical circuit is closed, without the application of any external voltage. Usually, a complete bulk intercalation takes 1-2 days period time, depending on the mass of the bulk crystals (2-5 mg) and the intrinsic kinetic of the process. After intercalation the crystal is disconnected from the circuit, rinsed thoroughly and left in a large volume of fresh MeOH or ACN for at least two hours. Then, the intercalated bulk crystal is placed in an oven at room temperature in a mild vacuum (~1 mbar) to accelerate the evaporation of the excess of solvent.

*Multiple flakes* – A bulk crystal is exfoliated using an adhesive tape (Nitto SPV224P) to obtain micro-sized and few-nm-thick flakes which are transferred on a thin film of Au(30 nm)/Ti(3 nm) film evaporated on glass. In the case of some vdW compounds ($\alpha$-$RuCl_3$, VOCl, CrOCl, CrSBr), the Au substrates are exposed to 10-15 s of air-plasma before the transfer of the flake, using a Plasma Pen Atmospheric Plasma System (PVA TePla), to improve the transfer yield.

Afterwards, the anodic metal is electrically connected to the Au film. In case of $In^0$ as the anode, it is sufficient to cut an $In^0$ piece and press it onto the surface of the substrate. In case of $Zn^0$ or $Mg^0$ as the anode, a piece of the metallic string is carefully sandpapered, cut in a ~ 25 $mm^2$ plate, and electrically connected to Au surface using $In^0$ pieces and a copper wire (Figure 3a). The system is immersed in an Eppendorf microcentrifuge tube containing the desired electrolyte (Figure 1e) for a certain amount of time (between 20 min. and 1h 30 min, see Supplementary Table 1). The substrate is finally rinsed with MeOH or ACN to remove the excess of the electrolyte.

*Single flakes* – Flakes are mechanically exfoliated from a bulk crystal using an adhesive tape (Nitto SPV224P) and transferred onto a Polydimethylsiloxane elastomer (PDMS). Homogeneous flakes with a typical thickness of 15-20 nm are selected and transferred onto prepatterned Au(15)/Ti(5) electrodes using a home-made delamination-stamping system. A piece of $In^0$ is pressed onto the pad of one contact or, in case of $Zn^0$ or $Mg^0$ as the anode, a fragment of this metal is cut and pressed onto the $In^0$ piece, which provides mechanical and electrical contact with the Au electrodes and the flake. Next, the electrochemical circuit is closed immersing the device in the electrolyte solution or, alternatively, by covering the device with a drop of the electrolyte solution. Typically, flakes are intercalated within seconds.

X-ray Diffractometry (XRD): XRD measurements were carried out using a Empyrean diffractometer (PANalytical) on bulk crystals and on exfoliated flakes supported on a Au(30-50)/Ti(3)/$SiO_2$ substrate. A copper cathode was used as X-ray source. Both the wavelengths $K\alpha_1$ (1.5406 Å) and $K\alpha_2$ (1.5443 Å) were employed to maximize the intensity of the diffracted beam.

Scanning Transmission Electron Microscopy. STEM studies were performed using a probe- aberration-corrected Thermo Fisher Scientific Titan Low Base microscope. This STEM instrument is equipped with a high-brightness gun (X-FEG) and an energy-dispersive X-ray spectrometer (EDS) (Ultim X-MaxN 100TLE detector, Oxford instruments). All measurements were performed at an electron energy of 300 keV. Convergence angle was 25 mrad and acceptance angle for high-angle annular dark-field (HAADF)-STEM imaging 48 mrad . STEM-EDS quantification was performed with the Aztec software (Oxford instruments).

A Thermo Fisher Scientific Helios650 focused ion beam (FIB) instrument has been employed for TEM lamella preparation.

Micro-Raman spectroscopy: Raman spectroscopy measurements of pristine and intercalated 2H-$TaS_2$ flakes were performed under ambient conditions with a Renishaw inVia Qontor equipped with a Nikon 100× objective (NA = 0.85) and a diffraction grating of 2400 lines $mm^{-1}$. A 532 nm laser was used with a power of ~ 1.7 mW for pristine and ~ 0.3 mW for the intercalated phase (~ 0.1 mW for few-layers flakes). Integration time was adjusted to enhance signal-to-noise ratio. In this case of the intensity maps of the $TMA^+$ intercalated 2H-$TaS_2$ device (see Supplementary Figure 16) and $TEA^+$ partially intercalated 2H-$TaS_2$ flake (see Figure 3e) the 532 nm laser power was set at ~ 1.7 mW and each spectrum was collected with an integration time of 15 s. Measurements were separated by 0.5 μm spacing.

Vibrating sample magnetometry (VSM): Magnetization vs. temperature or field H measurements were carried out using a physical properties measurement system (PPMS, Quantum Design) in vibrating sample magnetometer (VSM) mode.

Electrical transport measurements: The electrical measurements are recorded using a physical property measurement system (PPMS, Quantum Design) down to a temperature of 1.9 K. The resistance of pristine and intercalated flakes of 2H-$TaS_2$ and 2H-$NbSe_2$ are recorded via a four-points measurements, using a Keithley 6221 as a current source and a Keithley 2182 as a nanovoltmeter, operated in delta mode.

**Data availability**

All data in the main text or Supplementary Information are available from the corresponding author on reasonable request.

**Acknowledgements**

This work was supported under Projects PID2021-128004NB-C21, PID2021-122511OB-I00, PID2024-157558NB-C22, PID2024-157558NB-C21, and PID2024-155708OB-I00 funded by Spanish MICIU/AEI/10.13039/501100011033 and by ERDF/EU; and under the María de Maeztu Units of Excellence Programme (Grant CEX2020-001038-M). This work was supported by the FLAG-ERA grant MULTISPIN, via the Spanish MICIU/AEI with grant number PCI2021-122038-2A, and by the Diputación Foral de Gipuzkoa (QUANTUM, project no. 2024-QUAN-000014-01). It was also funded by the MCIN and by the European Union NextGenerationEU/PRTR-C17.I1, as well as by IKUR Strategy under the collaboration agreement between DONOSTIA INTERNATIONAL PHYSICS CENTER and NanoGune on behalf of the Department of Education of the Basque Government. B. M.-G. and M. G. acknowledge support from the "Ramón y Cajal" Programme by the Spanish MICIU/AEI and EU NextGenerationEU/PRTR (grant no. RYC2021-034836-I and RYC2021-031705-I). D.T. and C.A.-G. acknowledge the Spanish MICIU and ESF+ (grant no. PRE2020-092992 and PRE2022-104487).

Z.S. was supported by ERC-CZ program (project LL2101) from Ministry of Education Youth and Sports (MEYS) by the project Advanced Functional Nanorobots (reg. No. CZ.02.1.01/0.0/0.0/15_003/0000444 financed by the ERDF). V.M. was supported by project LUAUS23049 from Ministry of Education Youth and Sports (MEYS). I. R. thanks Ikerbasque and acknowledges support from Grant PID2023-151549NB-I00 and RED2022-134287-T, funded by MICIU and FEDER, EU, as well as Grant IT-1553-22 funded by the Basque Government. M.O. acknowledges the support from the UPV/EHU research project (EHU-N23/52). R. A. acknowledges funding from the Spanish MICIU with funding from European Union Next Generation EU (PRTR-C17.I1) promoted by the Government of Aragon and by the Spanish MICIU (PID2023-151080NB-I00 and CEX2023-001286-S), as well as from the Government of Aragon (DGA) through the project E13 23R. The studies were conducted at the Laboratorio de Microscopias Avanzadas (LMA), Universidad de Zaragoza, Spain.

Y.K.M. acknowledges support through a María Zambrano grant from the University of the Basque Country (Grant Ref. MAZAM21/12) funded by the Spanish MICIU the European Union NextGenerationEU.

A.M.-A. acknowledges support from the Basque Science Foundation for Science (Ikerbasque), POLYMAT, the University of the Basque Country, Diputación de Guipúzcoa, Gobierno Vasco (PIBA_2024_1_0030 and BERC programme) and Gobierno de España (Projects PID2021-124484OB-I00, PCI2022-132921, and María de Maeztu Excellence Unit CEX2023-001303-M funded by MICIU/AEI and European Union NextGenerationEU/PRTR). This project has received funding from the European Research Council (ERC) under the European Union's Horizon 2020 Research and Innovation Programme (Grant Agreement No. 722951). This Project has received funding from the European Union under the Horizon 2020 Research and Innovation Programme M-ERA.NET 2021 (SuperSuper). This work was funded by the European Union under the Horizon Europe grant 101046231 (FantastiCOF).

Technical and human support provided by SGIker of UPV/EHU is acknowledged.

**Author information**

*Contributions* D. T., M. O., M. G. conceived the study. D. T. and C. A. performed intercalations (bulk crystals and multi-flake) and the XRD measurements. D.T., C. A., D.M. and J.M.P. fabricated the devices for the intercalation of flakes and carried out the transport measurements. M.F. and R.A. performed the STEM characterization. D. T. and B. M.-G. performed the micro-Raman spectroscopy characterization. U. A., V. M. and Z. S. synthetized pristine bulk crystals of 2H-NbSe$_2$, 2H-TiS$_2$, 2H-SnS$_2$, FeOCl, VOCl, CrOCl, α-RuCl$_3$. Y.K.M. synthetized CoCp2I under the supervision of A.M-A. S. M.-V., E. C. synthetized pristine bulk crystals of CrSBr. F. M. S. and M. O. carried out the X-ray photoemission spectroscopy characterization. I. R. synthetized L- and D-*PrI. M. G., M. O., F. C. and L. E. H. supervised the work. D. T. and M. G. wrote the manuscript, with inputs from all co-authors.

*Corresponding authors* Correspondence to Marco Gobbi (marco.gobbi@ehu.eus)

**Ethics declarations**

*Competing interests* – The authors declare no competing interests.

**Supplementary information**

Supplementary Methods 1 – 2, Figs. 1–17, Tables 1–2.